\documentclass[prl,aps,showpacs,reprint,superscriptaddress,amsmath,amssymb,floatfix,10pt,longbibliography]{revtex4-1}
\usepackage{graphicx}
\usepackage{amsmath,amssymb,graphicx}
\usepackage[colorlinks=true, citecolor=blue, linkcolor=blue, urlcolor=blue]{hyperref}
\usepackage{array}
\usepackage{bm}

\begin{document}

%
\newcommand{\mat}[4]{\left(\begin{array}{cc}{#1}&{#2}\\[1mm]{#3}&{#4}
                     \end{array}\right)}
\newcommand{\matn}[4]{\left(\!\begin{array}{cc}{#1}\!&{#2}\\[1mm]{#3}\!&{#4}
                     \end{array}\!\right)}
\newcommand{\matnn}[4]{\left(\!\!
                      \begin{array}{cc}{#1}\!\!&{#2}\\[1mm]
                                       {#3}\!\!&{#4}
                      \end{array}\!\!\right)}
\newcommand{\mao}[4]{\begin{array}{cc}{#1}&{#2}\\[1mm]{#3}&{#4}
                     \end{array}}
\title{Revealing charge anisotropies in metal compounds \\via high-purity x-ray polarimetry}

\author{Lena Scherthan}
\affiliation{Fachbereich Physik, Technische Universit\"at Kaiserslautern, Erwin-Schr\"odinger Str. 46, 67663 Kaiserslautern, Germany}

\author{Juliusz A. Wolny}
\affiliation{Fachbereich Physik, Technische Universit\"at Kaiserslautern, Erwin-Schr\"odinger Str. 46, 67663 Kaiserslautern, Germany}

\author{Isabelle Faus}
\affiliation{Fachbereich Physik, Technische Universit\"at Kaiserslautern, Erwin-Schr\"odinger Str. 46, 67663 Kaiserslautern, Germany}

\author{Olaf Leupold}
\affiliation{Deutsches Elektronen-Synchrotron DESY, Notkestr. 85, 22607 Hamburg, Germany}

\author{Kai S. Schulze}
\affiliation{Helmholtz-Institut Jena, Fr\"obelstieg 3, 07743 Jena, Germany}
\affiliation{Institut f\"ur Optik und Quantenelektronik, Friedrich-Schiller-Universit\"at Jena, Max-Wien-Platz 1, 07743 Jena, Germany}

\author{Sebastian H\"ofer}
\affiliation{Helmholtz-Institut Jena, Fr\"obelstieg 3, 07743 Jena, Germany}
\affiliation{Institut f\"ur Optik und Quantenelektronik, Friedrich-Schiller-Universit\"at Jena, Max-Wien-Platz 1, 07743 Jena, Germany}

\author{Robert Loetzsch}
\affiliation{Helmholtz-Institut Jena, Fr\"obelstieg 3, 07743 Jena, Germany}
\affiliation{Institut f\"ur Optik und Quantenelektronik, Friedrich-Schiller-Universit\"at Jena, Max-Wien-Platz 1, 07743 Jena, Germany}

\author{Berit Marx-Glowna}
\affiliation{Helmholtz-Institut Jena, Fr\"obelstieg 3, 07743 Jena, Germany}
\affiliation{Institut f\"ur Optik und Quantenelektronik, Friedrich-Schiller-Universit\"at Jena, Max-Wien-Platz 1, 07743 Jena, Germany}

\author{Christopher E. Anson}
\affiliation{Institut f\"ur Anorganische Chemie, Karlsruher Institut f\"ur Technologie (KIT), Engesserstrasse 15, 76131 Karlsruhe, Germany}

\author{Annie K. Powell}
\affiliation{Institut f\"ur Anorganische Chemie, Karlsruher Institut f\"ur Technologie (KIT), Engesserstrasse 15, 76131 Karlsruhe, Germany}
\affiliation{Institut f\"ur Nanotechnologie, Karlsruher Institut f\"ur Technologie (KIT), Hermann-von-Helmholtz-Platz 1, 76344 Eggenstein-Leopoldshafen, Germany}

\author{Ingo Uschmann}
\affiliation{Helmholtz-Institut Jena, Fr\"obelstieg 3, 07743 Jena, Germany}
\affiliation{Institut f\"ur Optik und Quantenelektronik, Friedrich-Schiller-Universit\"at Jena, Max-Wien-Platz 1, 07743 Jena, Germany}

\author{Hans-Christian Wille}
\affiliation{Deutsches Elektronen-Synchrotron DESY, Notkestr. 85, 22607 Hamburg, Germany}

\author{Gerhard Paulus}
\affiliation{Helmholtz-Institut Jena, Fr\"obelstieg 3, 07743 Jena, Germany}
\affiliation{Institut f\"ur Optik und Quantenelektronik, Friedrich-Schiller-Universit\"at Jena, Max-Wien-Platz 1, 07743 Jena, Germany}

\author{Volker Sch\"unemann}
\affiliation{Fachbereich Physik, Technische Universit\"at Kaiserslautern, Erwin-Schr\"odinger Str. 46, 67663 Kaiserslautern, Germany}

\author{Ralf R\"ohlsberger}
\affiliation{Deutsches Elektronen-Synchrotron DESY, Notkestr. 85, 22607 Hamburg, Germany}
\affiliation{Helmholtz-Institut Jena, Fr\"obelstieg 3, 07743 Jena, Germany}
\affiliation{Institut f\"ur Optik und Quantenelektronik, Friedrich-Schiller-Universit\"at Jena, Max-Wien-Platz 1, 07743 Jena, Germany}

\vskip 0.25cm
\date{\today}

\begin{abstract}
Linear polarization analysis of hard x-rays is employed to probe electronic anisotropies in metal-containing complexes with very high selectivity. We use the pronounced linear dichroism of nuclear resonant x-ray scattering to determine electric field gradients in an iron(II) containing compound as they evolve during a temperature-dependent  high-spin/low-spin phase transition. This method constitutes a novel approach to analyze changes in the electronic structure of metal-containing molecules as function of external parameters or stimuli. The polarization selectivity of the technique allows us to monitor defect concentrations of electronic valence states across phase transitions. This opens new avenues to trace electronic changes and their precursors that are connected to structural and electronic dynamics in the class of metal compounds ranging from simple molecular solids to biological molecules. 
\end{abstract}

\maketitle

The knowledge of charge distributions and their anisotropies in molecules is crucial for understanding and precise modeling of molecular interactions and optical properties of chemical systems \cite{Kramer2014}. Powerful spectroscopic methods to investigate charge anisotropies in molecules are optical linear dichroism and birefringence that rely on probing the orientation dependence of the interaction of linearly polarized light with matter \cite{Norden2010}. For example, optical linear dichroism is applied to follow the orientation or reorientation of molecular systems resulting from external stimuli, like any reaction that involves changing the structure or length of a molecule or structural phase transitions in molecular solids \cite{Che1994,Rehault2011}. For a microscopic analysis of charge distributions in molecules, however, one needs to apply optical techniques capable of probing atomic length scales.

The availability of highly brilliant synchrotron radiation sources has facilitated the transfer of linear and circular dichroism to the regime of hard x-rays \cite{vanderLaan1986,Schuetz1987}. Today, x-ray linear and circular dichroism are very powerful techniques to probe charge and spin order in condensed matter with elemental specificity and site selectivity \cite{Lovesey1996,Stoehr2006}. Dichroic spectroscopies probe the imaginary part of the anisotropic index of refraction $n$ by comparing the incoherent absorption of two orthogonal polarization states. The same information is also contained in the real part of $n$ that leads to a rotation of the plane of polarization when two orthogonal polarization components experience different coherent phase shifts, an effect well known as optical birefringence. The analysis of birefringence in the x-ray regime, however, is much less common than x-ray dichroism, because efficient polarization analysis at energies of hard x-rays requires some significant technical effort. On the other hand, it opens a series of advantages compared to dichroism, because (a) it can be combined with momentum-resolving methods for high spatial resolution, (b) it does not require two subsequent measurements with different polarization states which allows for time-resolved measurements, e.g., in pump-probe schemes and (c) it promises to obtain significantly higher signal-to-noise ratios. 

Recently, high-purity polarimetry in the x-ray regime has been shown to be a very sensitive technique to detect tiny changes in the polarization state of x-rays caused by the interaction with an anisotropic or chiral, thus optically active medium \cite{Marx2013}. The technique relies on the application of x-ray linear Bragg polarizers in a 90$^{\circ}$ crossed setting with purities in the range of 10$^{-10}$ \cite{Schulze2018}. This method is most sensitively applied at atomic absorption edges \cite{Siddons1990,Hart1991,Schmitt2020} or nuclear resonances \cite{Toellner1995,Siddons1995,Roehlsberger1997,Alp2000,Heeg2013,Haber2016}.

\begin{figure}[t!]
\centering
\includegraphics[width=0.9\linewidth]{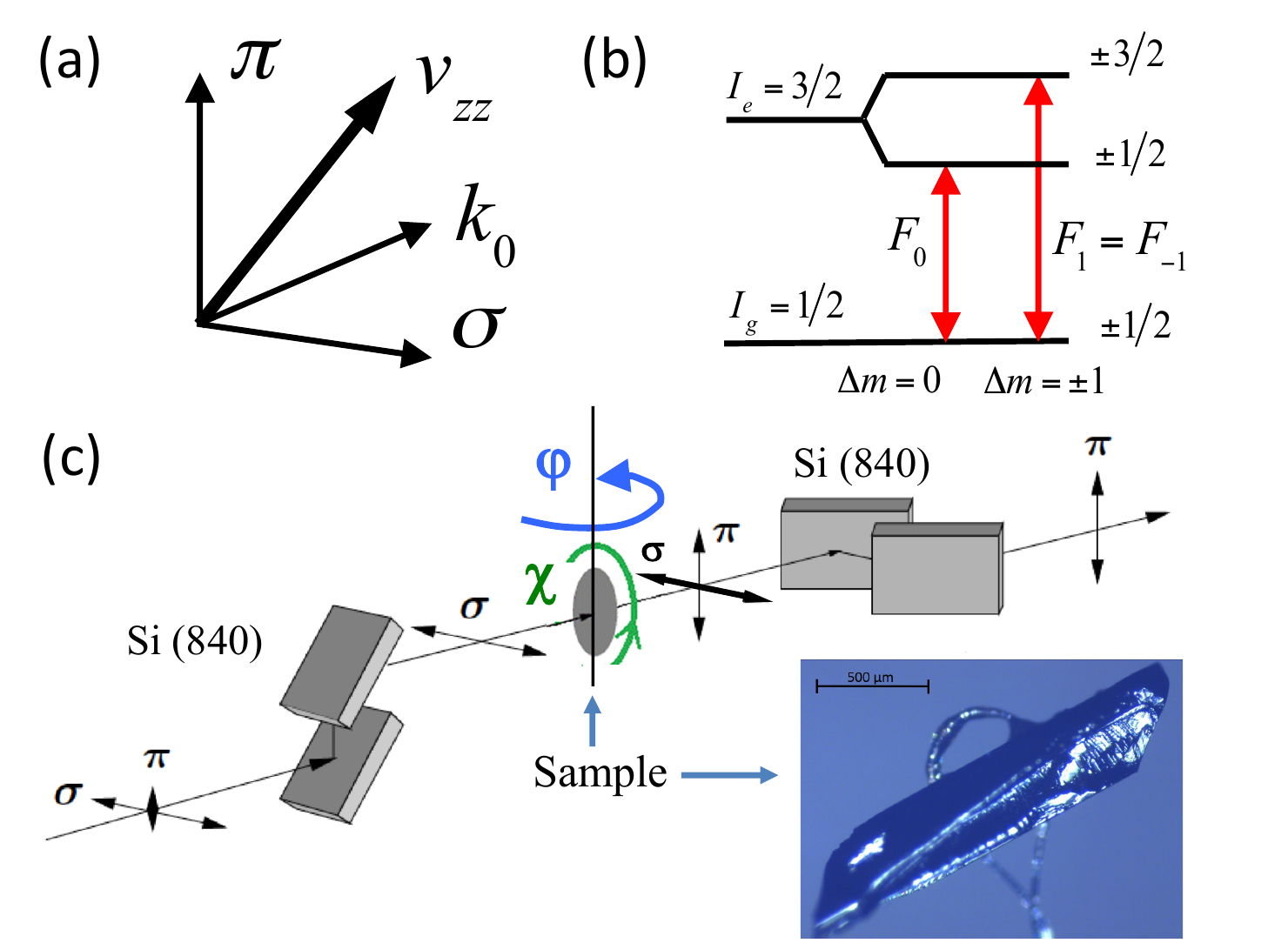}
\caption{(a) Sketch of the nuclear forward scattering (NFS) geometry, where $\bm{k}_0, \bm{\sigma}, \bm{\pi}$ and $\bm{v}_{zz}$ are unit vectors of the photon wavevector, the linear polarization basis, and the main component of the electric field gradient in the sample, respectively. (b) Nuclear level scheme of $^{57}$Fe for the case of a pure electric hyperfine interaction. (c) Experimental setup for polarization-resolved nuclear forward scattering (PR-NFS) experiments, consisting of two polarizing Si(840) channelcut crystals that are aligned in a 90$^{\circ}$ crossed setting to detect $\sigma \rightarrow \pi$ scattering from the sample located in between. In the experiment, the transmission through the polarimeter is monitored as function of the sample orientation, adjusted via the angles  $\varphi$ and $\chi$. The image at the bottom right shows a micrograph of the sample crystal.}
\label{Fig1}
\end{figure}

Here we employ polarization-resolved nuclear forward scattering (PR-NFS) of synchrotron radiation at the 14.4-keV nuclear resonance of $^{57}$Fe to probe atomic charge anisotropies in a metal compound. If the charge anisotropy in the system leads to an electric field gradient (EFG) at the $^{57}$Fe nucleus, nuclear states with a finite electrical quadrupole moment are subject to an electric hyperfine interaction. As a result, the excited nuclear state of $^{57}$Fe splits into two sublevels, shown in Fig.\,\ref{Fig1}b, which leads to a pronounced optical activity in the vicinity of the nuclear resonance. The nuclear resonant forward scattering amplitude $\mathbf{f}_r$ is given by \cite{Roehlsberger2004}

\begin{equation}
\mathbf{f}_r \sim F_1\,\mathbf{I} +
\matnn{(\bm{\pi}\cdot\bm{v}_{zz})^2 }
{ (\bm{\sigma}\cdot\bm{v}_{zz}) (\bm{\pi}\cdot\bm{v}_{zz})}
{ (\bm{\sigma}\cdot\bm{v}_{zz}) (\bm{\pi}\cdot\bm{v}_{zz})}
{(\bm{\sigma}\cdot\bm{v}_{zz})^2}[F_0 - F_1]
\end{equation}

where $\mathbf{I}$ is the $(2\times 2)$ unity matrix, $\bm{v}_{zz}$ is a unit vector along the main axis of the electric field gradient, $\sigma$ and $\pi$ are the linear polarization basis vectors (see Fig.\ref{Fig1}a) and $F_0, F_1$ describe Lorentzian functions that represent the resonant transitions between the nuclear ground state and the two levels of the excited nuclear state  with $\Delta m = 0$ and $\Delta m = \pm 1$, respectively. If the main axis of the EFG is not parallel or perpendicular to the incident linear $\sigma$ polarization, the off-diagonal elements of the matrix quantity $\mathbf{f}_r$ are non-zero and the polarization state of the transmitted radiation turns out to be elliptical. Thus, probing the $\pi$-component of the transmitted radiation in a setup with two crossed polarizers provides a sensitive measure of the orientation of the EFG relative to the plane of linear polarization of the incident light. This property is exploited here to probe the reorientation of electrical field gradients upon a high-spin (HS, S=2)/low-spin (LS, S=0) phase transition in an iron(II) containing spin crossover (SCO) compound \cite{Guetlich1994ie}. 

The spin state of iron(II) SCO compounds can be switched reversibly from the LS state to the HS state by variation of temperature, pressure or by irradiation with light \cite{Kahn1998,Guetlich2004,Halcrow2013}. More recently, the spin dependent charge transport properties of SCO molecules have also generated interest for their use in spintronic devices \cite{Gopakumar2012,Ruiz2014}. Indeed, a giant resistance change across the phase transition in SCO materials of more than 3000\,$\%$ has been reported theoretically \cite{Baadji2012}. Although not yet realized, it can be envisioned that the spin dependent charge and/or spin transfer properties of such devices may be tuned by variable concentrations of minority HS defect sites in a matrix of LS centers or vice versa. Therefore, it is of pivotal interest to have a spectral method at hand which is able to probe specifically spin dependent changes in the electronic anisotropy of minority spin species in SCO compounds. 

As a reference model for PR-NFS at the 14.4 keV nuclear resonance of $^{57}$Fe,  the SCO compound [Fe(PM-BiA)$_2$(NCS)$_2$] was chosen \cite{Letard2003}. This compound shows a thermally inducible, reversible spin state transition between the LS and the HS state in the range 150 - 230 K for powder samples. In the HS state, it exhibits a pure electric hyperfine interaction caused by a large electric field gradient (EFG) at the position of the Fe atom typical for high-spin octahedral Fe(II) complexes \cite{Letard2003,Guetlich1994ie}. A single crystal of [Fe(PM-BiA)$_2$(NCS)$_2$], enriched in $^{57}$Fe, was grown from dichloromethane/ethanol/diethyl ether.
The monoclinic phase of the synthesized single crystals, with four molecules in its unit cell (see Supplemental Material \cite{SuppMat}, Fig.\,S1) was confirmed by X-ray crystallographic measurements. This phase shows a gradual SCO behavior with an almost pure HS state at 220 K and above.

The experiments were carried out at beamline P01, PETRA III, DESY, Hamburg \cite{P01} in 60 bunch mode of operation, providing a bunch separation of 128 ns. The energy of the synchrotron radiation was tuned to the 14.4 keV resonance of $^{57}$Fe with a bandwidth of about 1 eV using a Si(111) double-crystal monochromator. The experimental setup is schematically shown in Fig.\,\ref{Fig1}c. For the PR-NFS experiments, the sample and its environment are located between two polarizing, 4-bounce Si(840) channelcut crystals that were cut with an asymmetry angle of -28.1$^{\circ}$ to optimally transmit the incident x-ray beam. The two channelcuts are aligned in crossed setting in order to detect nuclear resonant $\sigma \rightarrow \pi$ scattering. Due to the very high suppression of the non-resonant $\sigma$-polarized background to a level on the order of 10$^{-9}$, no high-resolution monochromator was needed to perform these experiments. The energetic bandpass of the (840) reflection at these crystals is 43\,meV with 75$\%$ peak transmission, constituting a significant energetic bandpass reduction to avoid radiation damage at the sample.

Orientation dependent NFS and PR-NFS experiments at various temperatures were performed with the sample mounted on a Eulerian Cradle for angular orientation, see Fig.\,\ref{Fig1}c, whereby the cooling of the sample down to temperatures in the range of 100 K was performed by the use of a N$_2$ cryogenic gas stream \cite{Rackwitz2014}. A series of NFS time spectra of the SCO compound were taken at room temperature without polarization analyzer for selected orientations of the sample to confirm the magnitude of the quadrupole splitting ($\Delta$E$_Q$ = 2.58 ($\pm$0.02) mm/s, asymmetry parameter $\eta = 0.3 (\pm 0.2)$ in the Fe(II) HS state (see Supplemental Material \cite{SuppMat}, Fig. S2).

\begin{figure}[t]
\centering
\includegraphics[width=0.85\linewidth]{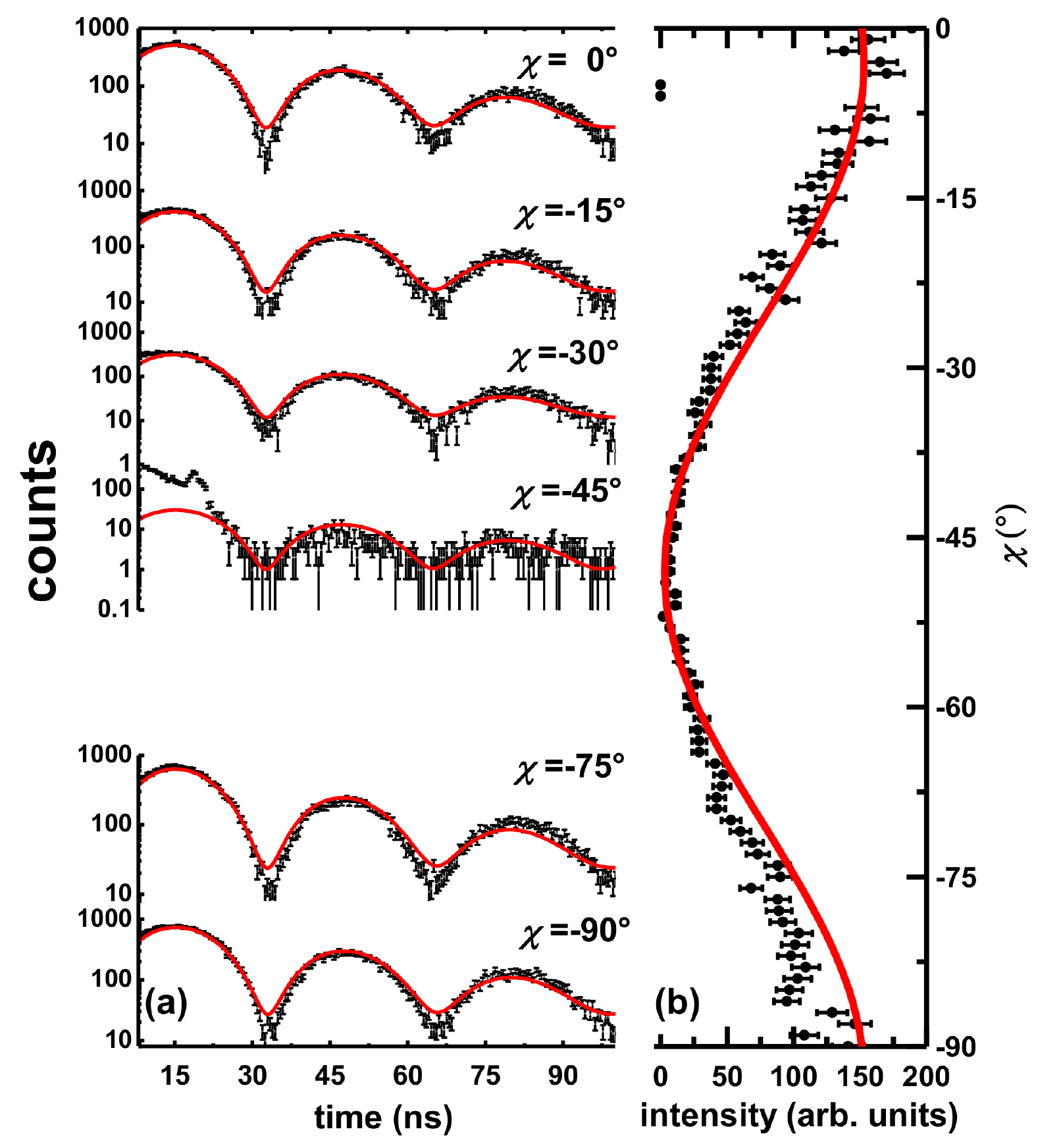}
\caption{(a) PR-NFS spectra recorded for selected orientations $\chi$ of the single crystal at $T$ = 220 K and $\varphi =0$ (black). The red lines show theoretical calculations of the scattered amplitude assuming a pure HS state with $\Delta$E$_Q$ = 2.64 ($\pm$0.15) mm/s and $\eta = 0.3\,(\pm 0.2)$.  (b) Corresponding $\chi$-PR-NFS-scan with theoretical calculation (red line) showing clearly the orientation dependence of the nuclear resonant $\sigma \rightarrow \pi$ scattering. Parameters are given in Table S1 in the Supplemental Material \cite{SuppMat}. The deviation between the simulation and the measured data for $\chi$ = -45$^{\circ}$ is due to a spurious bunch and electronic noise.}
\label{Fig2}
\end{figure}

High-purity polarimetry was then performed to determine the direction of the electric field gradients in the sample. A series of PR-NFS spectra taken at $T$ = 220\,K are shown in Fig.\,\ref{Fig2}a. For each orientation $\chi$, the data exhibit a temporal beat pattern resulting from a quadrupole splitting  of $\Delta$E$_Q$ = 2.64 ($\pm$0.15) mm/s. 

In addition to the time-dependent NFS and PR-NFS spectra, the time-integrated PR-NFS signal was recorded as a function of the angle $\chi$ while the angle $\varphi$ was kept fix (from now on referred to as $\chi$-PR-NFS-scans). In Fig.\,\ref{Fig2}b, such a $\chi$-PR-NFS-scan for $\varphi =0^{\circ}$ is shown. In accordance with the PR-NFS spectra, this $\chi$-PR-NFS scan exhibits a minimum at $\chi \sim -45^{\circ}$, which means that at this angle the main component of the EFG is either parallel or perpendicular to the incident linear $\sigma$-polarization. 
\begin{figure}[t!]
\centering
\includegraphics[width=\linewidth]{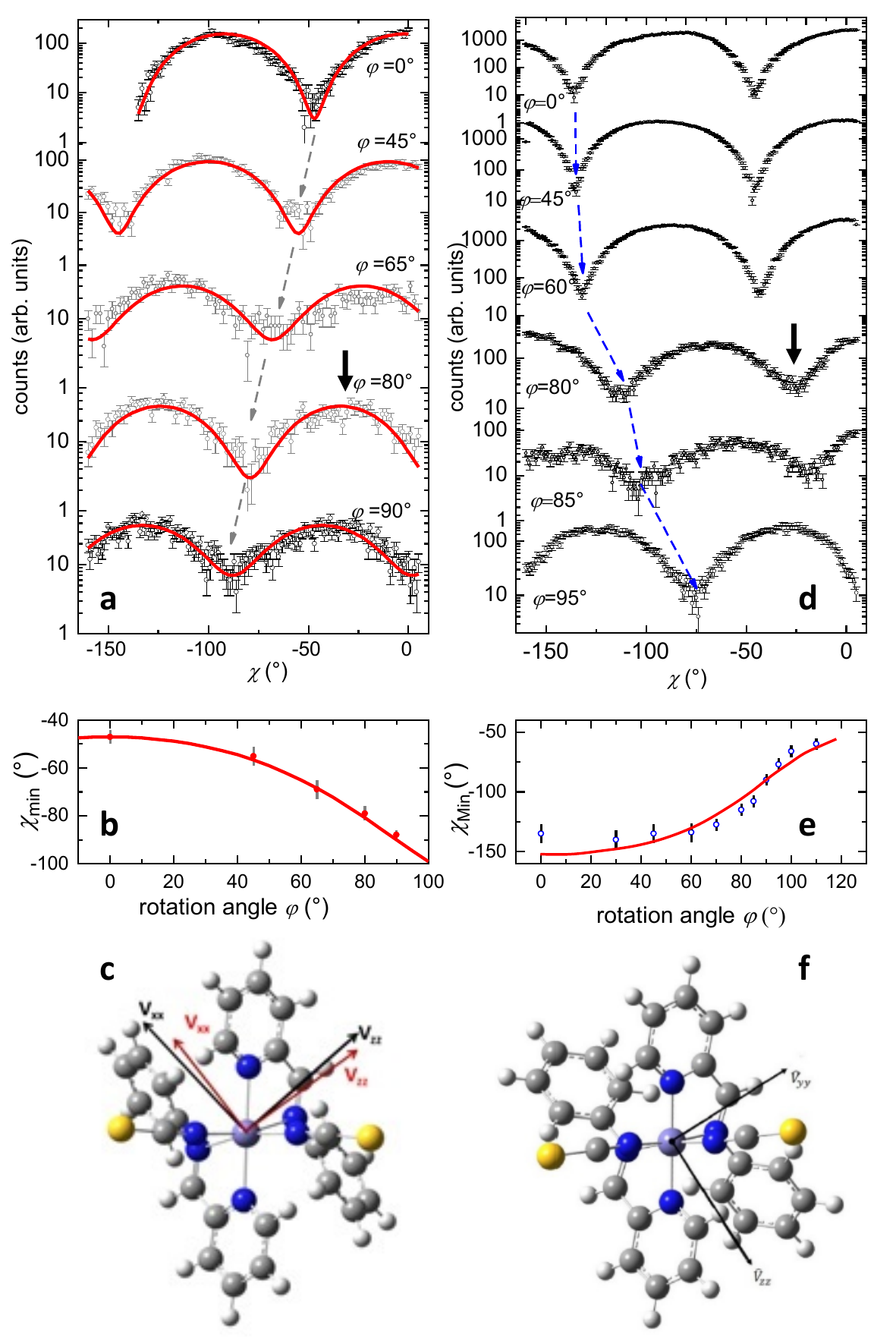}
\caption{(a,d): Nuclear resonant $\sigma - \pi$ scattering ($\chi$-PR-NFS-scans) taken at temperatures $T \geq$ 220 K (a) and $T$ = 120 K (d) for selected orientations $\varphi$. The temperature dependent change of the $\chi$-PR-NFS-scans proves the reorientation of the EFG upon the SCO transition from the HS to the LS state. The dashed lines are guides to the eye to trace the shift of two minima of the curves for the HS and the LS state, respectively. The figures below (b,e) display the shift of these minima as function of the rotation angle $\varphi$, starting at $\varphi$ = 0 . 
(c,f): Schematic representation of [Fe(PM-BiA)$_2$(NCS)$_2$] with the orientation of the EFG at the site of the Fe atom in the HS state (c) and the LS state (f) determined via a DFT calculation assuming C$_2$ symmetry (black vectors). The viewing direction, which corresponds to the direction of the $V_{yy}$ component of the HS EFG and to the $V_{xx}$ component of the LS EFG, is along the c-axis of the single crystal. For the HS state, the red arrows show the EFG orientation as determined from the experimental data, evidenced by the solid red lines in (a). The solid red lines in (b,e) are fits according to Eq.\,(S3) in the Supplemental Material \cite{SuppMat}, taking the EFG orientations shown in (c,f) as configuration at $\varphi = 0$. 
}
\label{Fig3}
\end{figure}
When the sample is oriented at either  $\chi \sim 0^{\circ}$ or $\chi \sim -90^{\circ}$, the off-diagonal elements of  $\mathbf{f}_r$ are maximized. Consequently, the time-integrated $\chi$-PR-NFS-scan has a maximum and the time-dependent PR-NFS spectrum shows a quantum beat pattern characteristic for a HS species. Thus, the $\chi$-PR-NFS-scans clearly demonstrate the sensitivity of the $\sigma \rightarrow \pi$ scattering process to the atomic charge anisotropy at the position of the $^{57}$Fe nucleus. 
Furthermore, these PR-NFS-scans, delivering a full orientation dependence, are less time consuming ($\sim$\,30 min per scan over a $\varphi$ range of 180$^{\circ}$ at RT) than taking NFS or PR-NFS time spectra for several orientations ($\sim$\,10 min per orientation $(\varphi, \chi)$ at a countrate of  $\sim$\, 500\,s$^{-1}$ at RT).
Therefore we recorded a series of $\chi$-PR-NFS-scans over a $\chi$-interval from 0$^{\circ}$  to -160$^{\circ}$ for different values of $\varphi$, shown in Figs.\,\ref{Fig3}a. The positions of the maxima/minima in the $\chi$-PR-NFS scans display a pronounced dependence on the sample orientation $\varphi$, shown in Figs.\,\ref{Fig3}b.

The detailed analysis of these data with the use of the dynamical theory of nuclear forward scattering \cite{Sturhahn2000,CONUSS} (solid red lines in Figs.\,\ref{Fig3}a,b) yields the EFG orientation of the HS state in the molecular coordinate system (Fig.\,\ref{Fig3}c) and reproduces all obtained NFS time spectra taken at room temperature (see Fig. S2, Supplemental Material \cite{SuppMat}). The EFG's main axis system of the molecular HS state determined in this way is depicted in Fig.\,\ref{Fig3}c together with a main axis system calculated by density functional calculations (DFT) assuming a C$_2$ symmetry \cite{DFT}.
Allowing a tolerance of 5$^{\circ}$ the theoretical calculation reproduces the experimentally determined EFG quite well.  

\begin{figure}[t!]
\centering
\includegraphics[width=\linewidth]{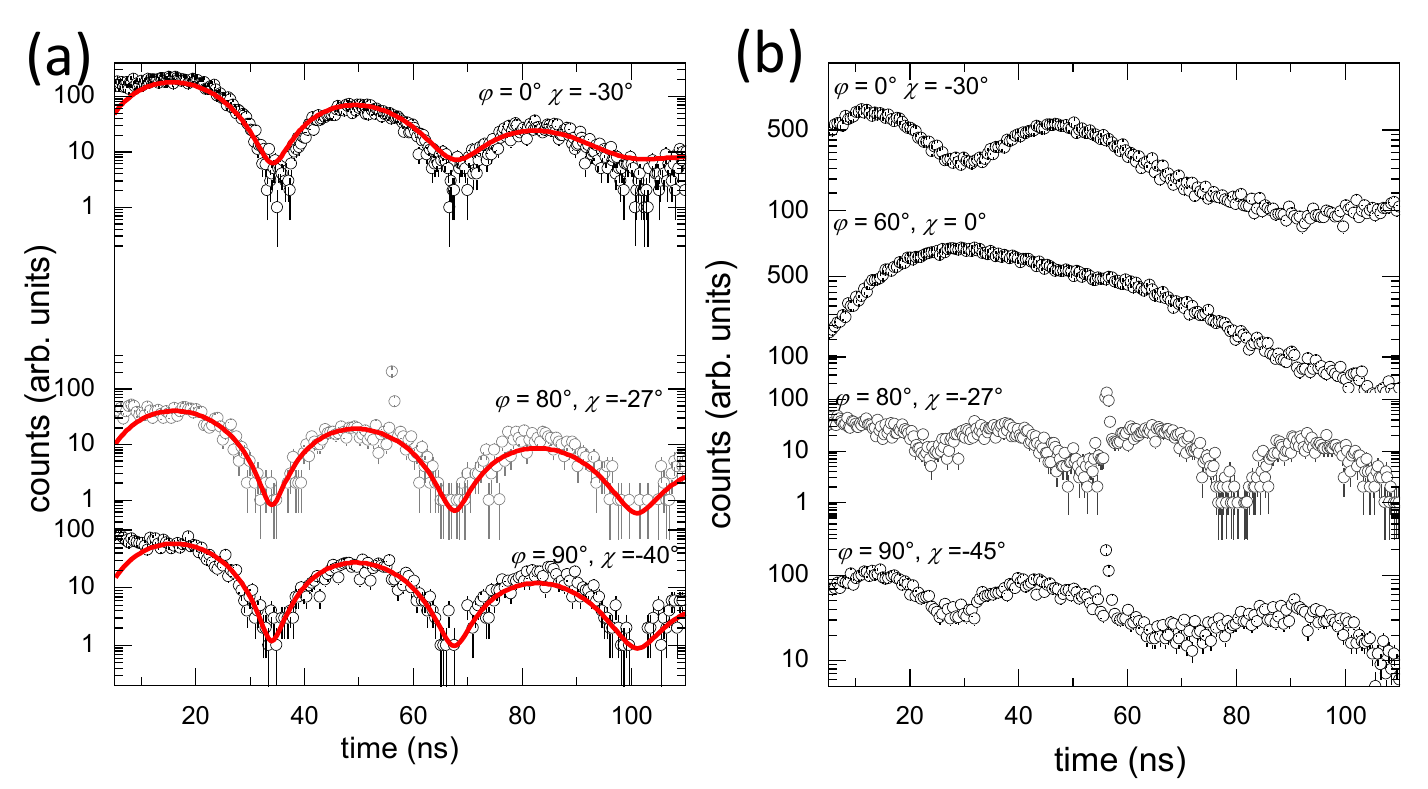}
\caption{PR-NFS beat patterns taken (a) at temperatures $T \geq$ 220 K (gray: room temperature, black: $T$ = 220 K) and (b) at $T$ = 120 K, for selected orientations  $(\varphi, \chi)$. While the data in (a) all show the quantum beat pattern characteristic for the quadrupole splitting $\Delta E_Q$ of the HS state, the beat pattern of the data taken at $T$ = 120 K is strongly orientation dependent, indicating a superposition of contributions from the LS state in the majority phase and the HS state in a 5$\%$ minority phase with different EFG orientations.}
\label{Fig4}
\end{figure}

At $T$ = 120 K the monoclinic phase of [Fe(PM-BiA)$_2$(NCS)$_2$] should be mostly in the LS state. In fact, orientation dependent NFS data obtained at $T$ = 120 K show that the majority of the iron centers are in their LS state which can be inferred from the low value of the quadrupole splitting ($\Delta E_Q$ = 0.63 ($\pm$0.02) mm/s, see Supplemental Material \cite{SuppMat}, Fig.\,S3). 
Fig.\,\ref{Fig3}d displays $\chi$-PR-NFS-scans obtained at $T$ =120 K in a $\chi$-interval from 0$^{\circ}$ to {-160}$^{\circ}$. 
Compared to the data in Fig.\,\ref{Fig3}a, the minima of the curves shift in opposite direction with increasing angle $\varphi$ as displayed in Fig.\,\ref{Fig3}e. This reflects a significant change of the EFG orientation as compared to the HS state that occurs during the SCO transition. Fig.\,\ref{Fig3}f shows the result of a DFT calculation of the EFG for the LS state. The red line in Fig.\,\ref{Fig3}e is a theoretical simulation considering the EFG of the DFT results as starting parameter for the analysis. While this curve qualitatively describes the measured data, deviations are clearly visible. An inspection of the PR-NFS time spectra at selected angular configurations $(\varphi, \chi)$ for the two temperature regimes, shown in Fig.\,\ref{Fig4}, points to the origin of these deviations, as discussed below.

In contrast to the high-temperature data in Fig.\,\ref{Fig4}a, the time spectra in Fig.\,\ref{Fig4}b taken at 120 K are strongly orientation dependent. While oscillations with a long period of $\sim$\,120 ns at ($\varphi = 60^{\circ}, \chi = 0^{\circ}$) reflect the temporal beat pattern of the quadrupole splitting of the LS state, the oscillations with the much shorter period of $\sim$\,30 ns at  ($\varphi = 80^{\circ}, \chi = -27^{\circ}$) carry the signnature of the HS state as in Fig.\ref{Fig4}a.  This observation suggests that at $T$ = 120 K there are still iron sites in the HS state which have not undergone a spin transition to the LS state. In fact, at this particular angular position, the $\chi$-PR-NFS scan in Fig.\,\ref{Fig3}d exhibits a deep minimum (black arrow), while the corresponding scan in Fig.\,\ref{Fig3}a has a maximum at this position (black arrow). This means that at ($\varphi = 80^{\circ}, \chi = -27^{\circ}$) the signal from the LS majority contribution is strongly suppressed by the polarimeter, thereby enabling the selective detection of the HS minority species that are still present at $T$ = 120 K. The beat patterns at the other orientations appear to be a superposition of long-period and short-period oscillation belonging to the LS and HS states. Since a signature from the HS state is clearly visible in the time spectra, it can only come from an ordered HS minority species with an EFG orientation that is different from that of the LS majority species. From the relative intensity of the HS signal in the time spectra, especially at ($\varphi = 80^{\circ}, \chi = -27^{\circ}$), we estimate a fraction of 0.05 of the HS state still present at 120 K. 

Without the crossed polarizers the signal from the majority LS state would be about a factor of (0.05)$^{-1}$ = 20 larger, rendering detection of the signal from the HS minority species very difficult if not impossible. Unfortunately, an accurate determination of the EFG orientation of the LS state from the experimental data is hampered by the large number of free parameters  at 120 K, as discussed in more detail in Sec.\,6 of the Supplemental Material \cite{SuppMat}.

In conclusion, polarization-resolved nuclear forward scattering (PR-NFS) was used here for the first time to investigate charge anisotropies via electric hyperfine interactions to gain insights into the ordering of minority phases. 
With the help of this novel technique it could be shown that during a first order SCO transition from a HS to a LS phase the electric field gradient tensor and thus the charge anisotropy of a HS center is not affected by the increasing amount of LS centers. The HS centers retain their charge anisotropy even when present as a minority phase of a few percent. 

Such studies are facilitated by the the use of synchrotron radiation (i.e. high photon flux in small, collimated beams) with its enormous brilliance and its intrinsic linear polarization that enables one to very efficiently probe electronic anisotropies in micron-sized single crystals. Single phases of materials can be selectively investigated that would not be accessible in macroscopic polycrystalline or powder samples as typically used in conventional M\"ossbauer spectroscopy. PR-NFS is applicable to any M\"ossbauer isotope that exhibits a sizeable nuclear quadrupole moment to determine the electric field gradient at the location of the nucleus. This technique opens new avenues to trace electronic changes that are connected to structural and electronic dynamics in almost all M\"ossbauer-active metal-containing compounds ranging from biological molecules to solid state systems. Combined with pump-probe techniques our approach opens new perspectives to reveal electronic dynamics of nonequilibrium states in molecular systems that are created, e.g., by impulsive external stimuli like temperature, pressure or electromagnetic and optical pulses: Monitoring the transmission of the polarimeter in a fixed angular setting $(\varphi, \chi)$ could reveal the emergence of electronic order in selected anisotropy directions that are transiently populated during dynamical processes. In general, our method has the potential to sensitively detect ordering phenomena at the onset of phase transitions, transient phenomena, or evolution of nonequilibrium states.

\begin{acknowledgments}

We acknowledge the support of the Helmholtz Association through project-oriented funds. 
Moreover, this work was supported by the German BMBF in the frame of the projects 05K16UKA and 05K13SJ1, and by the Deutsche Forschungsgemeinschaft (DFG) through TRR 173 - 268565370 (Project A04). J.A.W. and V.S. are grateful to Allianz f\"ur Hochleistungsrechnen Rheinland-Pfalz (AHRP) for providing CPU-time within the project TUK-SPINPLUSVIB. 
\end{acknowledgments}


\begin{thebibliography}{34}%
\makeatletter
\providecommand \@ifxundefined [1]{%
 \@ifx{#1\undefined}
}%
\providecommand \@ifnum [1]{%
 \ifnum #1\expandafter \@firstoftwo
 \else \expandafter \@secondoftwo
 \fi
}%
\providecommand \@ifx [1]{%
 \ifx #1\expandafter \@firstoftwo
 \else \expandafter \@secondoftwo
 \fi
}%
\providecommand \natexlab [1]{#1}%
\providecommand \enquote  [1]{``#1''}%
\providecommand \bibnamefont  [1]{#1}%
\providecommand \bibfnamefont [1]{#1}%
\providecommand \citenamefont [1]{#1}%
\providecommand \href@noop [0]{\@secondoftwo}%
\providecommand \href [0]{\begingroup \@sanitize@url \@href}%
\providecommand \@href[1]{\@@startlink{#1}\@@href}%
\providecommand \@@href[1]{\endgroup#1\@@endlink}%
\providecommand \@sanitize@url [0]{\catcode `\\12\catcode `\$12\catcode
  `\&12\catcode `\#12\catcode `\^12\catcode `\_12\catcode `\%12\relax}%
\providecommand \@@startlink[1]{}%
\providecommand \@@endlink[0]{}%
\providecommand \url  [0]{\begingroup\@sanitize@url \@url }%
\providecommand \@url [1]{\endgroup\@href {#1}{\urlprefix }}%
\providecommand \urlprefix  [0]{URL }%
\providecommand \Eprint [0]{\href }%
\providecommand \doibase [0]{http://dx.doi.org/}%
\providecommand \selectlanguage [0]{\@gobble}%
\providecommand \bibinfo  [0]{\@secondoftwo}%
\providecommand \bibfield  [0]{\@secondoftwo}%
\providecommand \translation [1]{[#1]}%
\providecommand \BibitemOpen [0]{}%
\providecommand \bibitemStop [0]{}%
\providecommand \bibitemNoStop [0]{.\EOS\space}%
\providecommand \EOS [0]{\spacefactor3000\relax}%
\providecommand \BibitemShut  [1]{\csname bibitem#1\endcsname}%
\let\auto@bib@innerbib\@empty
\bibitem [{\citenamefont {Kramer}\ \emph {et~al.}(2014)\citenamefont {Kramer},
  \citenamefont {Spinn},\ and\ \citenamefont {Liedl}}]{Kramer2014}%
  \BibitemOpen
  \bibfield  {author} {\bibinfo {author} {\bibfnamefont {C.}~\bibnamefont
  {Kramer}}, \bibinfo {author} {\bibfnamefont {A.}~\bibnamefont {Spinn}}, \
  and\ \bibinfo {author} {\bibfnamefont {K.~R.}\ \bibnamefont {Liedl}},\
  }\bibfield  {title} {\enquote {\bibinfo {title} {Charge anisotropy: Where
  atomic multipoles matter most},}\ }\href {\doibase 10.1021/ct5005565}
  {\bibfield  {journal} {\bibinfo  {journal} {Journal of Chemical Theory and
  Computation}\ }\textbf {\bibinfo {volume} {10}},\ \bibinfo {pages}
  {4488--4496} (\bibinfo {year} {2014})}\BibitemShut {NoStop}%
\bibitem [{\citenamefont {Nord\'en}\ \emph {et~al.}(2010)\citenamefont
  {Nord\'en}, \citenamefont {Rodger},\ and\ \citenamefont
  {Dafforn}}]{Norden2010}%
  \BibitemOpen
  \bibfield  {author} {\bibinfo {author} {\bibfnamefont {B.}~\bibnamefont
  {Nord\'en}}, \bibinfo {author} {\bibfnamefont {A.}~\bibnamefont {Rodger}}, \
  and\ \bibinfo {author} {\bibfnamefont {T.}~\bibnamefont {Dafforn}},\
  }\href@noop {} {\emph {\bibinfo {title} {Linear Dichroism and Circular
  Dichroism}}}\ (\bibinfo  {publisher} {The Royal Society of Chemistry},\
  \bibinfo {year} {2010})\BibitemShut {NoStop}%
\bibitem [{\citenamefont {Che}\ \emph {et~al.}(1994)\citenamefont {Che},
  \citenamefont {Shapiro}, \citenamefont {Esquerra},\ and\ \citenamefont
  {Kliger}}]{Che1994}%
  \BibitemOpen
  \bibfield  {author} {\bibinfo {author} {\bibfnamefont {D.}~\bibnamefont
  {Che}}, \bibinfo {author} {\bibfnamefont {D.~B.}\ \bibnamefont {Shapiro}},
  \bibinfo {author} {\bibfnamefont {R.~M.}\ \bibnamefont {Esquerra}}, \ and\
  \bibinfo {author} {\bibfnamefont {D.~S.}\ \bibnamefont {Kliger}},\ }\bibfield
   {title} {\enquote {\bibinfo {title} {Ultrasensitive time-resolved linear
  dichroism spectral measurements using near-crossed linear polarizers},}\
  }\href {\doibase 10.1016/0009-2614(94)00530-3} {\bibfield  {journal}
  {\bibinfo  {journal} {Chemical Physics Letters}\ }\textbf {\bibinfo {volume}
  {224}},\ \bibinfo {pages} {145 -- 154} (\bibinfo {year} {1994})}\BibitemShut
  {NoStop}%
\bibitem [{\citenamefont {R\'ehault}\ \emph {et~al.}(2011)\citenamefont
  {R\'ehault}, \citenamefont {Zanirato}, \citenamefont {Olivucci},\ and\
  \citenamefont {Helbing}}]{Rehault2011}%
  \BibitemOpen
  \bibfield  {author} {\bibinfo {author} {\bibfnamefont {J.}~\bibnamefont
  {R\'ehault}}, \bibinfo {author} {\bibfnamefont {V.}~\bibnamefont {Zanirato}},
  \bibinfo {author} {\bibfnamefont {M.}~\bibnamefont {Olivucci}}, \ and\
  \bibinfo {author} {\bibfnamefont {J.}~\bibnamefont {Helbing}},\ }\bibfield
  {title} {\enquote {\bibinfo {title} {Linear dichroism amplification: Adapting
  a long-known technique for ultrasensitive femtosecond ir spectroscopy},}\
  }\href {\doibase 10.1063/1.3572334} {\bibfield  {journal} {\bibinfo
  {journal} {The Journal of Chemical Physics}\ }\textbf {\bibinfo {volume}
  {134}},\ \bibinfo {pages} {124516} (\bibinfo {year} {2011})}\BibitemShut
  {NoStop}%
\bibitem [{\citenamefont {van~der Laan}\ \emph {et~al.}(1986)\citenamefont
  {van~der Laan}, \citenamefont {Thole}, \citenamefont {Sawatzky},
  \citenamefont {Goedkoop}, \citenamefont {Fuggle}, \citenamefont {Esteva},
  \citenamefont {Karnatak}, \citenamefont {Remeika},\ and\ \citenamefont
  {Dabkowska}}]{vanderLaan1986}%
  \BibitemOpen
  \bibfield  {author} {\bibinfo {author} {\bibfnamefont {G.}~\bibnamefont
  {van~der Laan}}, \bibinfo {author} {\bibfnamefont {B.~T.}\ \bibnamefont
  {Thole}}, \bibinfo {author} {\bibfnamefont {G.~A.}\ \bibnamefont {Sawatzky}},
  \bibinfo {author} {\bibfnamefont {J.~B.}\ \bibnamefont {Goedkoop}}, \bibinfo
  {author} {\bibfnamefont {J.~C.}\ \bibnamefont {Fuggle}}, \bibinfo {author}
  {\bibfnamefont {J.-M.}\ \bibnamefont {Esteva}}, \bibinfo {author}
  {\bibfnamefont {R.}~\bibnamefont {Karnatak}}, \bibinfo {author}
  {\bibfnamefont {J.~P.}\ \bibnamefont {Remeika}}, \ and\ \bibinfo {author}
  {\bibfnamefont {H.~A.}\ \bibnamefont {Dabkowska}},\ }\bibfield  {title}
  {\enquote {\bibinfo {title} {Experimental proof of magnetic x-ray
  dichroism},}\ }\href {\doibase 10.1103/PhysRevB.34.6529} {\bibfield
  {journal} {\bibinfo  {journal} {Phys. Rev. B}\ }\textbf {\bibinfo {volume}
  {34}},\ \bibinfo {pages} {6529--6531} (\bibinfo {year} {1986})}\BibitemShut
  {NoStop}%
\bibitem [{\citenamefont {Sch\"utz}\ \emph {et~al.}(1987)\citenamefont
  {Sch\"utz}, \citenamefont {Wagner}, \citenamefont {Wilhelm}, \citenamefont
  {Kienle}, \citenamefont {Zeller}, \citenamefont {Frahm},\ and\ \citenamefont
  {Materlik}}]{Schuetz1987}%
  \BibitemOpen
  \bibfield  {author} {\bibinfo {author} {\bibfnamefont {G.}~\bibnamefont
  {Sch\"utz}}, \bibinfo {author} {\bibfnamefont {W.}~\bibnamefont {Wagner}},
  \bibinfo {author} {\bibfnamefont {W.}~\bibnamefont {Wilhelm}}, \bibinfo
  {author} {\bibfnamefont {P.}~\bibnamefont {Kienle}}, \bibinfo {author}
  {\bibfnamefont {R.}~\bibnamefont {Zeller}}, \bibinfo {author} {\bibfnamefont
  {R.}~\bibnamefont {Frahm}}, \ and\ \bibinfo {author} {\bibfnamefont
  {G.}~\bibnamefont {Materlik}},\ }\bibfield  {title} {\enquote {\bibinfo
  {title} {Absorption of circularly polarized x rays in iron},}\ }\href
  {\doibase 10.1103/PhysRevLett.58.737} {\bibfield  {journal} {\bibinfo
  {journal} {Phys. Rev. Lett.}\ }\textbf {\bibinfo {volume} {58}},\ \bibinfo
  {pages} {737--740} (\bibinfo {year} {1987})}\BibitemShut {NoStop}%
\bibitem [{\citenamefont {Lovesey}\ and\ \citenamefont
  {Collins}(1996)}]{Lovesey1996}%
  \BibitemOpen
  \bibfield  {author} {\bibinfo {author} {\bibfnamefont {S.W.}\ \bibnamefont
  {Lovesey}}\ and\ \bibinfo {author} {\bibfnamefont {S.P.}\ \bibnamefont
  {Collins}},\ }\href@noop {} {\emph {\bibinfo {title} {X-ray Scattering and
  Absorption by Magnetic Materials}}},\ Oxford Series on Synchrotron Radiation\
  (\bibinfo  {publisher} {Clarendon Press},\ \bibinfo {year}
  {1996})\BibitemShut {NoStop}%
\bibitem [{\citenamefont {St\"ohr}\ and\ \citenamefont
  {Siegmann}(2006)}]{Stoehr2006}%
  \BibitemOpen
  \bibfield  {author} {\bibinfo {author} {\bibfnamefont {J.}~\bibnamefont
  {St\"ohr}}\ and\ \bibinfo {author} {\bibfnamefont {H.~C.}\ \bibnamefont
  {Siegmann}},\ }\href@noop {} {\emph {\bibinfo {title} {Magnetism}}}\
  (\bibinfo  {publisher} {Springer Berlin Heidelberg},\ \bibinfo {year}
  {2006})\BibitemShut {NoStop}%
\bibitem [{\citenamefont {Marx}\ \emph {et~al.}(2013)\citenamefont {Marx},
  \citenamefont {Schulze}, \citenamefont {Uschmann}, \citenamefont {K\"ampfer},
  \citenamefont {L\"otzsch}, \citenamefont {Wehrhan}, \citenamefont {Wagner},
  \citenamefont {Detlefs}, \citenamefont {Roth}, \citenamefont {H\"artwig},
  \citenamefont {F\"orster}, \citenamefont {St\"ohlker},\ and\ \citenamefont
  {Paulus}}]{Marx2013}%
  \BibitemOpen
  \bibfield  {author} {\bibinfo {author} {\bibfnamefont {B.}~\bibnamefont
  {Marx}}, \bibinfo {author} {\bibfnamefont {K.~S.}\ \bibnamefont {Schulze}},
  \bibinfo {author} {\bibfnamefont {I.}~\bibnamefont {Uschmann}}, \bibinfo
  {author} {\bibfnamefont {T.}~\bibnamefont {K\"ampfer}}, \bibinfo {author}
  {\bibfnamefont {R.}~\bibnamefont {L\"otzsch}}, \bibinfo {author}
  {\bibfnamefont {O.}~\bibnamefont {Wehrhan}}, \bibinfo {author} {\bibfnamefont
  {W.}~\bibnamefont {Wagner}}, \bibinfo {author} {\bibfnamefont
  {C.}~\bibnamefont {Detlefs}}, \bibinfo {author} {\bibfnamefont
  {T.}~\bibnamefont {Roth}}, \bibinfo {author} {\bibfnamefont {J.}~\bibnamefont
  {H\"artwig}}, \bibinfo {author} {\bibfnamefont {E.}~\bibnamefont
  {F\"orster}}, \bibinfo {author} {\bibfnamefont {T.}~\bibnamefont
  {St\"ohlker}}, \ and\ \bibinfo {author} {\bibfnamefont {G.~G.}\ \bibnamefont
  {Paulus}},\ }\bibfield  {title} {\enquote {\bibinfo {title} {High-precision
  x-ray polarimetry},}\ }\href {\doibase 10.1103/PhysRevLett.110.254801}
  {\bibfield  {journal} {\bibinfo  {journal} {Phys. Rev. Lett.}\ }\textbf
  {\bibinfo {volume} {110}},\ \bibinfo {pages} {254801} (\bibinfo {year}
  {2013})}\BibitemShut {NoStop}%
\bibitem [{\citenamefont {Schulze}(2018)}]{Schulze2018}%
  \BibitemOpen
  \bibfield  {author} {\bibinfo {author} {\bibfnamefont {K.~S.}\ \bibnamefont
  {Schulze}},\ }\bibfield  {title} {\enquote {\bibinfo {title} {Fundamental
  limitations of the polarization purity of x rays},}\ }\href {\doibase
  10.1063/1.5061807} {\bibfield  {journal} {\bibinfo  {journal} {APL
  Photonics}\ }\textbf {\bibinfo {volume} {3}},\ \bibinfo {pages} {126106}
  (\bibinfo {year} {2018})}\BibitemShut {NoStop}%
\bibitem [{\citenamefont {Siddons}\ \emph {et~al.}(1990)\citenamefont
  {Siddons}, \citenamefont {Hart}, \citenamefont {Amemiya},\ and\ \citenamefont
  {Hastings}}]{Siddons1990}%
  \BibitemOpen
  \bibfield  {author} {\bibinfo {author} {\bibfnamefont {D.~P.}\ \bibnamefont
  {Siddons}}, \bibinfo {author} {\bibfnamefont {M.}~\bibnamefont {Hart}},
  \bibinfo {author} {\bibfnamefont {Y.}~\bibnamefont {Amemiya}}, \ and\
  \bibinfo {author} {\bibfnamefont {J.~B.}\ \bibnamefont {Hastings}},\
  }\bibfield  {title} {\enquote {\bibinfo {title} {X-ray optical activity and
  the faraday effect in cobalt and its compounds},}\ }\href {\doibase
  10.1103/PhysRevLett.64.1967} {\bibfield  {journal} {\bibinfo  {journal}
  {Phys. Rev. Lett.}\ }\textbf {\bibinfo {volume} {64}},\ \bibinfo {pages}
  {1967--1970} (\bibinfo {year} {1990})}\BibitemShut {NoStop}%
\bibitem [{\citenamefont {Hart}\ \emph {et~al.}(1991)\citenamefont {Hart},
  \citenamefont {Siddons}, \citenamefont {Amemiya},\ and\ \citenamefont
  {Stojanoff}}]{Hart1991}%
  \BibitemOpen
  \bibfield  {author} {\bibinfo {author} {\bibfnamefont {M.}~\bibnamefont
  {Hart}}, \bibinfo {author} {\bibfnamefont {D.~P.}\ \bibnamefont {Siddons}},
  \bibinfo {author} {\bibfnamefont {Y.}~\bibnamefont {Amemiya}}, \ and\
  \bibinfo {author} {\bibfnamefont {V.}~\bibnamefont {Stojanoff}},\ }\bibfield
  {title} {\enquote {\bibinfo {title} {Tunable x?ray polarimeters for
  synchrotron radiation sources},}\ }\href {\doibase 10.1063/1.1142523}
  {\bibfield  {journal} {\bibinfo  {journal} {Review of Scientific
  Instruments}\ }\textbf {\bibinfo {volume} {62}},\ \bibinfo {pages}
  {2540--2544} (\bibinfo {year} {1991})}\BibitemShut {NoStop}%
\bibitem [{\citenamefont {Schmitt}\ \emph {et~al.}(2020)\citenamefont
  {Schmitt}, \citenamefont {Joly}, \citenamefont {Schulze}, \citenamefont
  {Marx-Glowna}, \citenamefont {Uschmann}, \citenamefont {Grabiger},
  \citenamefont {Bernhardt}, \citenamefont {Loetzsch}, \citenamefont {Juhin},
  \citenamefont {Debray}, \citenamefont {Wille}, \citenamefont {Yavas},
  \citenamefont {Paulus},\ and\ \citenamefont {R\"ohlsberger}}]{Schmitt2020}%
  \BibitemOpen
  \bibfield  {author} {\bibinfo {author} {\bibfnamefont {A.}~\bibnamefont
  {Schmitt}}, \bibinfo {author} {\bibfnamefont {Y.}~\bibnamefont {Joly}},
  \bibinfo {author} {\bibfnamefont {K.~S.}\ \bibnamefont {Schulze}}, \bibinfo
  {author} {\bibfnamefont {B.}~\bibnamefont {Marx-Glowna}}, \bibinfo {author}
  {\bibfnamefont {I.}~\bibnamefont {Uschmann}}, \bibinfo {author}
  {\bibfnamefont {B.}~\bibnamefont {Grabiger}}, \bibinfo {author}
  {\bibfnamefont {H.}~\bibnamefont {Bernhardt}}, \bibinfo {author}
  {\bibfnamefont {R.}~\bibnamefont {Loetzsch}}, \bibinfo {author}
  {\bibfnamefont {A.}~\bibnamefont {Juhin}}, \bibinfo {author} {\bibfnamefont
  {J.}~\bibnamefont {Debray}}, \bibinfo {author} {\bibfnamefont {H.-C.}\
  \bibnamefont {Wille}}, \bibinfo {author} {\bibfnamefont {H.}~\bibnamefont
  {Yavas}}, \bibinfo {author} {\bibfnamefont {G.~G.}\ \bibnamefont {Paulus}}, \
  and\ \bibinfo {author} {\bibfnamefont {R.}~\bibnamefont {R\"ohlsberger}},\
  }\href@noop {} {\enquote {\bibinfo {title} {Disentangling {X}-ray dichroism
  and birefringence via high-purity polarimetry},}\ } (\bibinfo {year}
  {2020}),\ \Eprint {http://arxiv.org/abs/2003.00849} {arXiv:2003.00849
  [physics.ins-det]} \BibitemShut {NoStop}%
\bibitem [{\citenamefont {Toellner}\ \emph {et~al.}(1995)\citenamefont
  {Toellner}, \citenamefont {Alp}, \citenamefont {Sturhahn}, \citenamefont
  {Mooney}, \citenamefont {Zhang}, \citenamefont {Ando}, \citenamefont {Yoda},\
  and\ \citenamefont {Kikuta}}]{Toellner1995}%
  \BibitemOpen
  \bibfield  {author} {\bibinfo {author} {\bibfnamefont {T.~S.}\ \bibnamefont
  {Toellner}}, \bibinfo {author} {\bibfnamefont {E.~E.}\ \bibnamefont {Alp}},
  \bibinfo {author} {\bibfnamefont {W.}~\bibnamefont {Sturhahn}}, \bibinfo
  {author} {\bibfnamefont {T.~M.}\ \bibnamefont {Mooney}}, \bibinfo {author}
  {\bibfnamefont {X.}~\bibnamefont {Zhang}}, \bibinfo {author} {\bibfnamefont
  {M.}~\bibnamefont {Ando}}, \bibinfo {author} {\bibfnamefont {Y.}~\bibnamefont
  {Yoda}}, \ and\ \bibinfo {author} {\bibfnamefont {S.}~\bibnamefont
  {Kikuta}},\ }\bibfield  {title} {\enquote {\bibinfo {title}
  {Polarizer/analyzer filter for nuclear resonant scattering of synchrotron
  radiation},}\ }\href {\doibase 10.1063/1.114764} {\bibfield  {journal}
  {\bibinfo  {journal} {Applied Physics Letters}\ }\textbf {\bibinfo {volume}
  {67}},\ \bibinfo {pages} {1993} (\bibinfo {year} {1995})}\BibitemShut
  {NoStop}%
\bibitem [{\citenamefont {Siddons}\ \emph {et~al.}(1995)\citenamefont
  {Siddons}, \citenamefont {Hastings}, \citenamefont {Bergmann}, \citenamefont
  {Sette},\ and\ \citenamefont {Krisch}}]{Siddons1995}%
  \BibitemOpen
  \bibfield  {author} {\bibinfo {author} {\bibfnamefont {D.P.}\ \bibnamefont
  {Siddons}}, \bibinfo {author} {\bibfnamefont {J.B.}\ \bibnamefont
  {Hastings}}, \bibinfo {author} {\bibfnamefont {U.}~\bibnamefont {Bergmann}},
  \bibinfo {author} {\bibfnamefont {F.}~\bibnamefont {Sette}}, \ and\ \bibinfo
  {author} {\bibfnamefont {M.}~\bibnamefont {Krisch}},\ }\bibfield  {title}
  {\enquote {\bibinfo {title} {M{\"o}ssbauer spectroscopy using synchrotron
  radiation: overcoming detector limitations},}\ }\href {\doibase
  http://dx.doi.org/10.1016/0168-583X(95)00654-0} {\bibfield  {journal}
  {\bibinfo  {journal} {Nuclear Instruments and Methods in Physics Research
  Section B: Beam Interactions with Materials and Atoms}\ }\textbf {\bibinfo
  {volume} {103}},\ \bibinfo {pages} {371 -- 375} (\bibinfo {year}
  {1995})}\BibitemShut {NoStop}%
\bibitem [{\citenamefont {R{\"o}hlsberger}\ \emph {et~al.}(1997)\citenamefont
  {R{\"o}hlsberger}, \citenamefont {Gerdau}, \citenamefont {R{\"u}ffer},
  \citenamefont {Sturhahn}, \citenamefont {Toellner}, \citenamefont
  {Chumakov},\ and\ \citenamefont {Alp}}]{Roehlsberger1997}%
  \BibitemOpen
  \bibfield  {author} {\bibinfo {author} {\bibfnamefont {R.}~\bibnamefont
  {R{\"o}hlsberger}}, \bibinfo {author} {\bibfnamefont {E.}~\bibnamefont
  {Gerdau}}, \bibinfo {author} {\bibfnamefont {R.}~\bibnamefont {R{\"u}ffer}},
  \bibinfo {author} {\bibfnamefont {W.}~\bibnamefont {Sturhahn}}, \bibinfo
  {author} {\bibfnamefont {T.S.}\ \bibnamefont {Toellner}}, \bibinfo {author}
  {\bibfnamefont {A.I.}\ \bibnamefont {Chumakov}}, \ and\ \bibinfo {author}
  {\bibfnamefont {E.E.}\ \bibnamefont {Alp}},\ }\bibfield  {title} {\enquote
  {\bibinfo {title} {X-ray optics for ?ev-resolved spectroscopy},}\ }\href
  {\doibase http://dx.doi.org/10.1016/S0168-9002(97)00710-9} {\bibfield
  {journal} {\bibinfo  {journal} {Nuclear Instruments and Methods in Physics
  Research Section A: Accelerators, Spectrometers, Detectors and Associated
  Equipment}\ }\textbf {\bibinfo {volume} {394}},\ \bibinfo {pages} {251 --
  255} (\bibinfo {year} {1997})}\BibitemShut {NoStop}%
\bibitem [{\citenamefont {Alp}\ \emph {et~al.}(2000)\citenamefont {Alp},
  \citenamefont {Sturhahn},\ and\ \citenamefont {Toellner}}]{Alp2000}%
  \BibitemOpen
  \bibfield  {author} {\bibinfo {author} {\bibfnamefont {E.E.}\ \bibnamefont
  {Alp}}, \bibinfo {author} {\bibfnamefont {W.}~\bibnamefont {Sturhahn}}, \
  and\ \bibinfo {author} {\bibfnamefont {T.S.}\ \bibnamefont {Toellner}},\
  }\bibfield  {title} {\enquote {\bibinfo {title} {Polarizer--analyzer
  optics},}\ }\href {\doibase 10.1023/A:1012673301869} {\bibfield  {journal}
  {\bibinfo  {journal} {Hyperfine Interactions}\ }\textbf {\bibinfo {volume}
  {125}},\ \bibinfo {pages} {45--68} (\bibinfo {year} {2000})}\BibitemShut
  {NoStop}%
\bibitem [{\citenamefont {Heeg}\ \emph {et~al.}(2013)\citenamefont {Heeg},
  \citenamefont {Wille}, \citenamefont {Schlage}, \citenamefont {Guryeva},
  \citenamefont {Schumacher}, \citenamefont {Uschmann}, \citenamefont
  {Schulze}, \citenamefont {Marx}, \citenamefont {K\"ampfer}, \citenamefont
  {Paulus}, \citenamefont {R\"ohlsberger},\ and\ \citenamefont
  {Evers}}]{Heeg2013}%
  \BibitemOpen
  \bibfield  {author} {\bibinfo {author} {\bibfnamefont {K.~P.}\ \bibnamefont
  {Heeg}}, \bibinfo {author} {\bibfnamefont {H.-C.}\ \bibnamefont {Wille}},
  \bibinfo {author} {\bibfnamefont {K.}~\bibnamefont {Schlage}}, \bibinfo
  {author} {\bibfnamefont {T.}~\bibnamefont {Guryeva}}, \bibinfo {author}
  {\bibfnamefont {D.}~\bibnamefont {Schumacher}}, \bibinfo {author}
  {\bibfnamefont {I.}~\bibnamefont {Uschmann}}, \bibinfo {author}
  {\bibfnamefont {K.~S.}\ \bibnamefont {Schulze}}, \bibinfo {author}
  {\bibfnamefont {B.}~\bibnamefont {Marx}}, \bibinfo {author} {\bibfnamefont
  {T.}~\bibnamefont {K\"ampfer}}, \bibinfo {author} {\bibfnamefont {G.~G.}\
  \bibnamefont {Paulus}}, \bibinfo {author} {\bibfnamefont {R.}~\bibnamefont
  {R\"ohlsberger}}, \ and\ \bibinfo {author} {\bibfnamefont {J.}~\bibnamefont
  {Evers}},\ }\bibfield  {title} {\enquote {\bibinfo {title} {Vacuum-assisted
  generation and control of atomic coherences at x-ray energies},}\ }\href
  {\doibase 10.1103/PhysRevLett.111.073601} {\bibfield  {journal} {\bibinfo
  {journal} {Phys. Rev. Lett.}\ }\textbf {\bibinfo {volume} {111}},\ \bibinfo
  {pages} {073601} (\bibinfo {year} {2013})}\BibitemShut {NoStop}%
\bibitem [{\citenamefont {Haber}\ \emph {et~al.}(2016)\citenamefont {Haber},
  \citenamefont {Schulze}, \citenamefont {Schlage}, \citenamefont {Loetzsch},
  \citenamefont {Bocklage}, \citenamefont {Gurieva}, \citenamefont {Bernhardt},
  \citenamefont {Wille}, \citenamefont {R{\"u}ffer}, \citenamefont {Uschmann},
  \citenamefont {Paulus},\ and\ \citenamefont {R{\"o}hlsberger}}]{Haber2016}%
  \BibitemOpen
  \bibfield  {author} {\bibinfo {author} {\bibfnamefont {J.}~\bibnamefont
  {Haber}}, \bibinfo {author} {\bibfnamefont {K.~S.}\ \bibnamefont {Schulze}},
  \bibinfo {author} {\bibfnamefont {K.}~\bibnamefont {Schlage}}, \bibinfo
  {author} {\bibfnamefont {R.}~\bibnamefont {Loetzsch}}, \bibinfo {author}
  {\bibfnamefont {L.}~\bibnamefont {Bocklage}}, \bibinfo {author}
  {\bibfnamefont {T.}~\bibnamefont {Gurieva}}, \bibinfo {author} {\bibfnamefont
  {H.}~\bibnamefont {Bernhardt}}, \bibinfo {author} {\bibfnamefont {H.-C.}\
  \bibnamefont {Wille}}, \bibinfo {author} {\bibfnamefont {R.}~\bibnamefont
  {R{\"u}ffer}}, \bibinfo {author} {\bibfnamefont {I.}~\bibnamefont
  {Uschmann}}, \bibinfo {author} {\bibfnamefont {G.~G.}\ \bibnamefont
  {Paulus}}, \ and\ \bibinfo {author} {\bibfnamefont {R.}~\bibnamefont
  {R{\"o}hlsberger}},\ }\bibfield  {title} {\enquote {\bibinfo {title}
  {Collective strong coupling of x-rays and nuclei in a nuclear optical
  lattice},}\ }\href {http://dx.doi.org/10.1038/nphoton.2016.77} {\bibfield
  {journal} {\bibinfo  {journal} {Nat. Photon.}\ }\textbf {\bibinfo {volume}
  {10}},\ \bibinfo {pages} {445--449} (\bibinfo {year} {2016})}\BibitemShut
  {NoStop}%
\bibitem [{\citenamefont {R\"ohlsberger}(2004)}]{Roehlsberger2004}%
  \BibitemOpen
  \bibfield  {author} {\bibinfo {author} {\bibfnamefont {R.}~\bibnamefont
  {R\"ohlsberger}},\ }\href@noop {} {\emph {\bibinfo {title} {Nuclear Condensed
  Matter Physics with Synchrotron Radiation}}},\ \bibinfo {series} {Springer
  Tracts in Modern Physics}, Vol.\ \bibinfo {volume} {208}\ (\bibinfo
  {publisher} {Springer-Verlag Berlin Heidelberg},\ \bibinfo {year}
  {2004})\BibitemShut {NoStop}%
\bibitem [{\citenamefont {G\"utlich}\ \emph {et~al.}(1994)\citenamefont
  {G\"utlich}, \citenamefont {Hauser},\ and\ \citenamefont
  {Spiering}}]{Guetlich1994ie}%
  \BibitemOpen
  \bibfield  {author} {\bibinfo {author} {\bibfnamefont {Philipp}\ \bibnamefont
  {G\"utlich}}, \bibinfo {author} {\bibfnamefont {Andreas}\ \bibnamefont
  {Hauser}}, \ and\ \bibinfo {author} {\bibfnamefont {Hartmut}\ \bibnamefont
  {Spiering}},\ }\bibfield  {title} {\enquote {\bibinfo {title} {Thermal and
  optical switching of iron(ii) complexes},}\ }\href {\doibase
  10.1002/anie.199420241} {\bibfield  {journal} {\bibinfo  {journal}
  {Angewandte Chemie International Edition}\ }\textbf {\bibinfo {volume}
  {33}},\ \bibinfo {pages} {2024--2054} (\bibinfo {year} {1994})}\BibitemShut
  {NoStop}%
\bibitem [{\citenamefont {Kahn}\ and\ \citenamefont
  {Martinez}(1998)}]{Kahn1998}%
  \BibitemOpen
  \bibfield  {author} {\bibinfo {author} {\bibfnamefont {O.}~\bibnamefont
  {Kahn}}\ and\ \bibinfo {author} {\bibfnamefont {C.~Jay}\ \bibnamefont
  {Martinez}},\ }\bibfield  {title} {\enquote {\bibinfo {title}
  {Spin-transition polymers: From molecular materials toward memory devices},}\
  }\href {\doibase 10.1126/science.279.5347.44} {\bibfield  {journal} {\bibinfo
   {journal} {Science}\ }\textbf {\bibinfo {volume} {279}},\ \bibinfo {pages}
  {44--48} (\bibinfo {year} {1998})}\BibitemShut {NoStop}%
\bibitem [{\citenamefont {G{\"u}tlich}\ and\ \citenamefont
  {Goodwin}(2004)}]{Guetlich2004}%
  \BibitemOpen
  \bibfield  {author} {\bibinfo {author} {\bibfnamefont {P.}~\bibnamefont
  {G{\"u}tlich}}\ and\ \bibinfo {author} {\bibfnamefont {H.~A.}\ \bibnamefont
  {Goodwin}},\ }\enquote {\bibinfo {title} {Spin crossover---an overall
  perspective},}\ in\ \href {\doibase 10.1007/b13527} {\emph {\bibinfo
  {booktitle} {Spin Crossover in Transition Metal Compounds I}}},\ \bibinfo
  {editor} {edited by\ \bibinfo {editor} {\bibfnamefont {P.}~\bibnamefont
  {G{\"u}tlich}}\ and\ \bibinfo {editor} {\bibfnamefont {H.A.}\ \bibnamefont
  {Goodwin}}}\ (\bibinfo  {publisher} {Springer Berlin Heidelberg},\ \bibinfo
  {address} {Berlin, Heidelberg},\ \bibinfo {year} {2004})\ pp.\ \bibinfo
  {pages} {1--47}\BibitemShut {NoStop}%
\bibitem [{\citenamefont {Halcrow}(2013)}]{Halcrow2013}%
  \BibitemOpen
  \bibfield  {author} {\bibinfo {author} {\bibfnamefont {M.A.}\ \bibnamefont
  {Halcrow}},\ }\href {https://books.google.de/books?id=BC8OmFkLWLwC} {\emph
  {\bibinfo {title} {Spin-Crossover Materials: Properties and Applications}}}\
  (\bibinfo  {publisher} {Wiley},\ \bibinfo {year} {2013})\BibitemShut
  {NoStop}%
\bibitem [{\citenamefont {Gopakumar}\ \emph {et~al.}(2012)\citenamefont
  {Gopakumar}, \citenamefont {Matino}, \citenamefont {Naggert}, \citenamefont
  {Bannwarth}, \citenamefont {Tuczek},\ and\ \citenamefont
  {Berndt}}]{Gopakumar2012}%
  \BibitemOpen
  \bibfield  {author} {\bibinfo {author} {\bibfnamefont {T.~G.}\ \bibnamefont
  {Gopakumar}}, \bibinfo {author} {\bibfnamefont {F.}~\bibnamefont {Matino}},
  \bibinfo {author} {\bibfnamefont {H.}~\bibnamefont {Naggert}}, \bibinfo
  {author} {\bibfnamefont {A.}~\bibnamefont {Bannwarth}}, \bibinfo {author}
  {\bibfnamefont {F.}~\bibnamefont {Tuczek}}, \ and\ \bibinfo {author}
  {\bibfnamefont {R.}~\bibnamefont {Berndt}},\ }\bibfield  {title} {\enquote
  {\bibinfo {title} {Electron-induced spin crossover of single molecules in a
  bilayer on gold},}\ }\href {\doibase 10.1002/anie.201201203} {\bibfield
  {journal} {\bibinfo  {journal} {Angewandte Chemie International Edition}\
  }\textbf {\bibinfo {volume} {51}},\ \bibinfo {pages} {6262--6266} (\bibinfo
  {year} {2012})}\BibitemShut {NoStop}%
\bibitem [{\citenamefont {Ruiz}(2014)}]{Ruiz2014}%
  \BibitemOpen
  \bibfield  {author} {\bibinfo {author} {\bibfnamefont {Eliseo}\ \bibnamefont
  {Ruiz}},\ }\bibfield  {title} {\enquote {\bibinfo {title} {Charge transport
  properties of spin crossover systems},}\ }\href {\doibase 10.1039/C3CP54028F}
  {\bibfield  {journal} {\bibinfo  {journal} {Phys. Chem. Chem. Phys.}\
  }\textbf {\bibinfo {volume} {16}},\ \bibinfo {pages} {14--22} (\bibinfo
  {year} {2014})}\BibitemShut {NoStop}%
\bibitem [{\citenamefont {Baadji}\ and\ \citenamefont
  {Sanvito}(2012)}]{Baadji2012}%
  \BibitemOpen
  \bibfield  {author} {\bibinfo {author} {\bibfnamefont {N.}~\bibnamefont
  {Baadji}}\ and\ \bibinfo {author} {\bibfnamefont {S.}~\bibnamefont
  {Sanvito}},\ }\bibfield  {title} {\enquote {\bibinfo {title} {Giant
  resistance change across the phase transition in spin-crossover molecules},}\
  }\href {\doibase 10.1103/PhysRevLett.108.217201} {\bibfield  {journal}
  {\bibinfo  {journal} {Phys. Rev. Lett.}\ }\textbf {\bibinfo {volume} {108}},\
  \bibinfo {pages} {217201} (\bibinfo {year} {2012})}\BibitemShut {NoStop}%
\bibitem [{\citenamefont {L{\'e}tard}\ \emph {et~al.}(2003)\citenamefont
  {L{\'e}tard}, \citenamefont {Chastanet}, \citenamefont {Nguyen},
  \citenamefont {Marc{\'e}n}, \citenamefont {Marchivie}, \citenamefont
  {Guionneau}, \citenamefont {Chasseau},\ and\ \citenamefont
  {G{\"u}tlich}}]{Letard2003}%
  \BibitemOpen
  \bibfield  {author} {\bibinfo {author} {\bibfnamefont {J.-F.}\ \bibnamefont
  {L{\'e}tard}}, \bibinfo {author} {\bibfnamefont {G.}~\bibnamefont
  {Chastanet}}, \bibinfo {author} {\bibfnamefont {O.}~\bibnamefont {Nguyen}},
  \bibinfo {author} {\bibfnamefont {S.}~\bibnamefont {Marc{\'e}n}}, \bibinfo
  {author} {\bibfnamefont {M.}~\bibnamefont {Marchivie}}, \bibinfo {author}
  {\bibfnamefont {P.}~\bibnamefont {Guionneau}}, \bibinfo {author}
  {\bibfnamefont {D.}~\bibnamefont {Chasseau}}, \ and\ \bibinfo {author}
  {\bibfnamefont {P.}~\bibnamefont {G{\"u}tlich}},\ }\bibfield  {title}
  {\enquote {\bibinfo {title} {Spin crossover properties of the
  [{Fe(PM-BiA)$_2$(NCS)$_2$}] complex -- {Phases I and II}},}\ }\href {\doibase
  10.1007/s00706-002-0537-0} {\bibfield  {journal} {\bibinfo  {journal}
  {Monatshefte f{\"u}r Chemie / Chemical Monthly}\ }\textbf {\bibinfo {volume}
  {134}},\ \bibinfo {pages} {165--182} (\bibinfo {year} {2003})}\BibitemShut
  {NoStop}%
\bibitem [{Sup()}]{SuppMat}%
  \BibitemOpen
  \href@noop {} {}\bibinfo {note} {See Supplemental Material for details on
  sample alignment etc, which includes Refs.
  \cite{g09,Toellner1995,Siddons1999,Roehlsberger2004,Letard2003}}\BibitemShut
  {NoStop}%
\bibitem [{P01()}]{P01}%
  \BibitemOpen
  \href@noop {} {}\bibinfo {note} {{h}ttp://photon-science.desy.de/facilities/
  \\ petra$\_$iii/beamlines/p01$\_$dynamics/index$\_$eng.html}\BibitemShut
  {NoStop}%
\bibitem [{\citenamefont {Rackwitz}\ \emph {et~al.}(2014)\citenamefont
  {Rackwitz}, \citenamefont {Faus}, \citenamefont {Schmitz}, \citenamefont
  {Kelm}, \citenamefont {Kr{\"u}ger}, \citenamefont {Andersson}, \citenamefont
  {Hersleth}, \citenamefont {Achterhold}, \citenamefont {Schlage},
  \citenamefont {Wille}, \citenamefont {Sch{\"u}nemann},\ and\ \citenamefont
  {Wolny}}]{Rackwitz2014}%
  \BibitemOpen
  \bibfield  {author} {\bibinfo {author} {\bibfnamefont {S.}~\bibnamefont
  {Rackwitz}}, \bibinfo {author} {\bibfnamefont {I.}~\bibnamefont {Faus}},
  \bibinfo {author} {\bibfnamefont {M.}~\bibnamefont {Schmitz}}, \bibinfo
  {author} {\bibfnamefont {H.}~\bibnamefont {Kelm}}, \bibinfo {author}
  {\bibfnamefont {H.-J.}\ \bibnamefont {Kr{\"u}ger}}, \bibinfo {author}
  {\bibfnamefont {K.~K.}\ \bibnamefont {Andersson}}, \bibinfo {author}
  {\bibfnamefont {H.-P.}\ \bibnamefont {Hersleth}}, \bibinfo {author}
  {\bibfnamefont {K.}~\bibnamefont {Achterhold}}, \bibinfo {author}
  {\bibfnamefont {K.}~\bibnamefont {Schlage}}, \bibinfo {author} {\bibfnamefont
  {H.-C.}\ \bibnamefont {Wille}}, \bibinfo {author} {\bibfnamefont
  {V.}~\bibnamefont {Sch{\"u}nemann}}, \ and\ \bibinfo {author} {\bibfnamefont
  {J.~A.}\ \bibnamefont {Wolny}},\ }\bibfield  {title} {\enquote {\bibinfo
  {title} {A new sample environment for cryogenic nuclear resonance scattering
  experiments on single crystals and microsamples at {P01, PETRA III}},}\
  }\href {\doibase 10.1007/s10751-013-0981-8} {\bibfield  {journal} {\bibinfo
  {journal} {Hyperfine Interactions}\ }\textbf {\bibinfo {volume} {226}},\
  \bibinfo {pages} {673--678} (\bibinfo {year} {2014})}\BibitemShut {NoStop}%
\bibitem [{\citenamefont {Sturhahn}(2000)}]{Sturhahn2000}%
  \BibitemOpen
  \bibfield  {author} {\bibinfo {author} {\bibfnamefont {W.}~\bibnamefont
  {Sturhahn}},\ }\bibfield  {title} {\enquote {\bibinfo {title} {{CONUSS and
  PHOENIX}: Evaluation of nuclear resonant scattering data},}\ }\href {\doibase
  10.1023/A:1012681503686} {\bibfield  {journal} {\bibinfo  {journal}
  {Hyperfine Interactions}\ }\textbf {\bibinfo {volume} {125}},\ \bibinfo
  {pages} {149--172} (\bibinfo {year} {2000})}\BibitemShut {NoStop}%
\bibitem [{CON()}]{CONUSS}%
  \BibitemOpen
  \href@noop {} {}\bibinfo {note} {CONUSS is free software. It can be obtained
  via the software site http://www.nrixs.com.}\BibitemShut {Stop}%
\bibitem [{DFT()}]{DFT}%
  \BibitemOpen
  \href@noop {} {}\bibinfo {note} {DFT calculations were performed using
  Gaussian09 \cite{g09} with the functional B3LYP and the basis CEP-31G. We
  used a model molecule (shown in Fig.\,\ref{Fig3}) with one of the rings of
  biphenyl replaced with hydrogen in order to approximate the molecule with
  C$_2$ symmetry (compare with Fig.\,S1 in the Supplemental Material
  \cite{SuppMat}).}\BibitemShut {Stop}%
\end{thebibliography}

%

\end{document}